\documentclass[twocolumn,twocolappendix]{aastex701}

\usepackage{amsmath}
\usepackage{amssymb}
\usepackage{graphicx}
\usepackage{xcolor}

\graphicspath{{./}{figure/}}

\shorttitle{Gravitational Waves Trace Disk Reorientation}
\shortauthors{Xia \& Yuan}

\definecolor{revisionviolet}{RGB}{128,0,128}

\begin{document}

\title{The Geometric Imprint of a Tilted Accretion Disk in Gravitational Waves}

\makeatletter
\affiliation{Department of Physics, College of Sciences, Shanghai University,
99 Shangda Road, Shanghai 200444, China}
\move@AF
\let\AF@grp\@empty
\makeatother

\author[orcid=0009-0004-8669-2411,gname=Jing-Ze,sname=Xia]{Jing-Ze Xia}
\affiliation{Tsung-Dao Lee Institute, Shanghai Jiao Tong University,
1 Lisuo Road, Shanghai 201210, PR China}
\email{jackxia@sjtu.edu.cn}

\author[gname=Chen,sname=Yuan]{Chen Yuan}
\affiliation{Department of Physics, College of Sciences, Shanghai University,
99 Shangda Road, Shanghai 200444, China}
\email{yuanchen@shu.edu.cn}
\correspondingauthor{Chen Yuan}
\email{yuanchen@shu.edu.cn}

\begin{abstract}
Gravitational wave astronomy has opened a new window onto black holes and the dynamics of strong-field gravity, while accretion flows offer a complementary and natural laboratory for probing strong-field physics.
In this work, we show that the gravitational wave emission from an initially tilted accretion disk carries a distinct imprint of its orientation. We perform three dimensional general relativistic magnetohydrodynamic simulations and evaluate the gravitational waves through Teukolsky equation. For an aligned disk, the emission is concentrated in low-$|m|$ modes, where $m$ is the azimuthal quantum number. For the initially tilted disk, higher-$|m|$ modes make a larger relative contribution. As the tilted flow subsequently aligns with the black hole spin, the contribution of $|m|\ge2$ modes decreases, showing strong correlation with the tilt angle. This indicates that the gravitational waves generated by the black hole environment carries geometric information about the surrounding matter. Our results establish a route towards using multimode gravitational radiation to probe the geometry and dynamical evolution of accretion flows in the strong-field region.
\end{abstract}


\section{Introduction}
\label{sec:introduction}

The first direct detection of gravitational waves (GWs) from the merger of two black holes (BHs) inaugurated the era of GW astronomy \citep{Abbott2016GW150914}. The observation of the binary-neutron-star merger GW170817 extended this new window to multimessenger astrophysics
\citep{Abbott2017BNS}.
Beyond the study of compact binary populations, GWs can also serve as a direct probe of strong-field physics close to BH horizons \citep{LVK2026GW250114}. 
In addition to being messengers of compact-object dynamics, GWs can be generated by matter in the strong gravitational field of a BH, offering a way to probe the BH environment. A natural astrophysical setting for such environmental GW emission is a rotating BH surrounded by an accretion disk, in which matter and magnetic fields evolve in the near-horizon gravitational field.
Motivated by this possibility, GW emission from accretion disks has been explored in several source regimes. Early studies considered hyper-accretion onto newly formed BHs \citep{ArayaGochez2004}. General relativistic simulations of massive, self-gravitating BH disks projected that their signals could be detected by future ground-based and space-based observatories \citep{Wessel2021}. Cooled collapsar disks provide another promising channel, in which the signal may be within reach of third-generation ground-based detectors \citep{Gottlieb2024}, and simulations have linked GW signatures to shock oscillations in these systems \citep{Fernandez:2025qbq}. In a different regime, \citet{Yuan2025} evaluated the stochastic GW background from accretion disks and estimated that GWs sourced by turbulence driven by the magnetorotational instability (MRI) could fall in the microhertz band. More recently, fully relativistic magnetohydrodynamic simulations of self-gravitating BH disks found persistent nonaxisymmetric structure and associated GW emission, supporting these systems as multimessenger sources \citep{Ruiz2026}. Together, these studies show that GWs generated by accretion disks can probe strong-field dynamics and may be observable with future detectors. Motivated by these developments, we ask whether GWs generated by the accretion environment encode information about the spacetime geometry of that environment.

Disk orientation is one property that can shape both the accretion flow and its GW emission\citep{Ruiz2026}.  The angular momentum of an accretion flow around an individual BH need not coincide with the BH spin axis\citep{maccarone2002,kingpringle2006}.  Maser observations of the Circinus galaxy resolve a warped disk associated with a wide-angle outflow \citep{Greenhill2003Circinus}, while long-term X-ray observations of Her X-1 support a precessing, tilted inner disk \citep{Kosec2023HerX1}. More recently, correlated X-ray and radio variability in a tidal disruption event has been interpreted as evidence for Lense–Thirring-driven disk–jet co-precession \citep{Wang:2025oaq}. These measurements establish disk warping and precession as observable features of accreting systems, rather than merely idealized initial conditions \citep{Greenhill2003Circinus,Kosec2023HerX1}.  Frame dragging induces Lense--Thirring precession, while competition between warp communication and viscous dissipation can produce Bardeen--Petterson alignment \citep{Bardeen1975,ScheuerFeiler1996,King2005,Nealon2015}. 

Relativistic simulations predict further departures from planar, stationary accretion when the disk is geometrically thick or strongly tilted\citep{Chatterjee2025}. Hydrodynamic and general relativistic magnetohydrodynamic (GRMHD) calculations have demonstrated global precession, radial warps, shocks, and, in some regimes, disk tearing \citep{Fragile2005,Nixon2012,White2019,Liska2018}.  In magnetically arrested or otherwise strongly magnetized flows, large-scale magnetic flux can thread the horizon and extract BH rotational energy through the Blandford--Znajek mechanism \citep{BlandfordZnajek1977,Tchekhovskoy2011MAD}.  Three dimensional GRMHD simulations further show that magnetic stresses can collimate precessing outflows and exchange angular momentum among the BH, disk, and jet, leading to magneto-spin alignment or magnetically driven retrograde precession \citep{Liska2018,Chatterjee2025,Jiang2025Retrograde}.

These observations and simulations establish disk orientation as a
physically important ingredient, but the corresponding features of GWs from a turbulent tilted disk remain unclear. In this paper, we explore the features of GWs from an initially tilted disk around a rotating BH. We carefully study the relationship between the GW signals and the geometric inclination of the accretion disk. This paper is organized as follows.  Section~\ref{sec:model} describes the matched disk models, the time-domain perturbation calculation, and the angular diagnostics.  Section~\ref{sec:results} presents the multimode signature associated with disk alignment.  Section~\ref{sec:scaling} describes the physical scaling, while Section~\ref{sec:discussion} considers the interpretation and limitations of the proposed diagnostic.  Section~\ref{sec:conclusions} summarizes our main conclusions.

\section{Physical model and observables}
\label{sec:model}

\subsection{GRMHD setup}


To investigate the influence of disk geometry on GW energy mode distribution, we have performed three dimensional (3D) GRMHD simulations  with initially aligned disk (tilt angle $0^\circ$) and initially tilted disk (tilt angle $75^\circ$) using \texttt{AthenaK}\citep{Stone2020AthenaPP,Stone2024AthenaK}. The simulations evolve magnetized Fishbone--Moncrief tori \citep{Fishbone1976} on a fixed Kerr spacetime with
dimensionless spin $\chi\equiv a/M=0.94$, where $a$ is the Kerr spin parameter and $M$ is the BH mass. The torus inner edge and pressure maximum are $r_{\rm in}=10M$ and $r_{\rm max}=20M$. To generate a seed poloidal magnetic field, we adopt a single weak magnetic loop vector potential \citep{Tchekhovskoy2011MAD,narayan2012},
\begin{equation}
 A_\phi\propto\max\left[
 \frac{\rho}{\rho_{\max}}
 \left(\frac{r\sin\vartheta}{10 r_{\rm g}}\right)^3
 e^{-r/(400 r_{\rm g})}-0.2,\,0\right],
 \label{eq:aphi}
\end{equation}
where $\rho$ is the rest-mass density, $\rho_{\max}$ is its initial maximum, and $\vartheta$ denotes the polar angle in the tilted torus coordinate frame (aligned with the disk's angular momentum axis, rather than the computational grid coordinates. The field is normalized to $\max(p_{\rm gas})/\max(p_{\rm mag})=100$, where $p_{\rm mag}=b^2/2$ and $b^2=b^\mu b_\mu$ is the comoving magnetic energy density invariant. In the torus, we add a 2\% internal-energy perturbation to seed the magneto-rotational instability (MRI) \citep{balbus1991}. Here we use the units $G=c=1$. 

The aligned and tilted simulations use the same equation of state with adiabatic index $\Gamma=4/3$, grid, torus, magnetic field configuration. The only control parameter is the initial angle, $0^\circ$ or $75^\circ$, between the disk angular momentum and the BH spin.  The numerical setup, near-horizon accretion rate, normalized magnetic flux, and MRI quality factors are described in Appendices~\ref{app:accretion-flux} and~\ref{app:mri}.

\begin{figure*}[t]
\centering
\includegraphics[width=\textwidth]{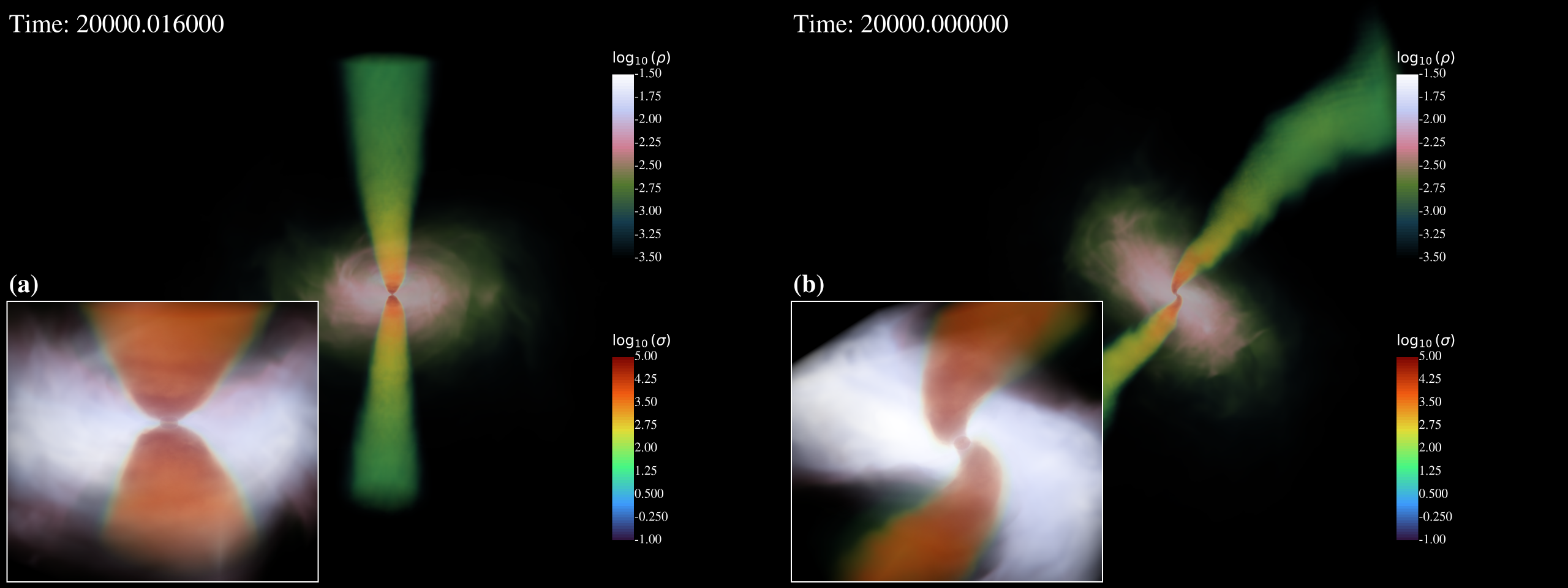}
\caption{Matched accretion-flow morphology near $t=2\times10^4M$.  Panel (a)
shows the aligned model and panel (b) the initially $75^\circ$
model, whose disk-body tilt has decreased to $34.0^\circ$.
The volume rendering combines rest-mass density, $\log_{10}\rho$, with the
magnetization $\log_{10}\sigma$, where $\sigma=b^2/\rho$.  The tilted model
shows a correspondingly inclined disk and magnetized outflow, whereas the
aligned model remains symmetric about the BH spin axis.
}
\label{fig:morphology}
\end{figure*}

\subsection{Teukolsky formalism}
\label{sec:teukolsky}

Firstly, we extract the total stress--energy tensor of the whole spacetime,
\begin{equation}
 T_{\mu\nu}=(\rho+\rho\epsilon+P+b^2)u_\mu u_\nu
 +\left(P+\frac{b^2}{2}\right)g_{\mu\nu}-b_\mu b_\nu .
 \label{eq:stress}
\end{equation}
Here $\rho$, $\epsilon$, and $P$ are the rest-mass density, specific internal energy and pressure of the fluid. $u^\mu$ is the fluid four-velocity and $b^\mu$ is the magnetic four-vector, with $b^2=b^\mu b_\mu$. We absorb the usual factor of $4\pi$ into $b^\mu$. We then numerically evolve each azimuthal source component with the time-domain Teukolsky equation \citep{Teukolsky1973,Yuan2025}.  The master variable is
\begin{equation}
 \Psi=(r-ia\cos\theta)^4\Psi_4
 =\sum_m r^3e^{im\phi}\Phi_m(t,r_*,\theta),
 \label{eq:master}
\end{equation}
with $\Psi_4=(\ddot h_+-i\ddot h_\times)/2$. Here $\Psi$ is the Teukolsky master variable, $\Psi_4$ is the Newman--Penrose curvature scalar, $m$ is the azimuthal quantum number and $r_*$ is the tortoise coordinate. The quantities $h_+$ and $h_\times$ are the two polarizations of GWs.

\subsection{Disk tilt and GW modes}

We define the orientation of the disk from its total angular momentum,
following the stress-energy-based construction used for tilted GRMHD disks
\citep{Fragile2005,Liska2018,Jiang2025Retrograde}.  Here $\mathcal V_{\rm d}$ denote the main disk-body volume, restricted to $2M\le r\le20M$, where $r$ is the radial coordinate and $M$ is the BH mass. The integrated four-momentum and orbital angular-momentum tensor are
\begin{align}
 S_{\rm d}^{\sigma}
 &=\int_{\mathcal V_{\rm d}} T^{\sigma 0}\,d^3x, \\
 L_{\rm d}^{\mu\nu}
 &=\int_{\mathcal V_{\rm d}}
 \left(x^\mu T^{\nu 0}-x^\nu T^{\mu 0}\right)d^3x .
 \label{eq:disk-four-momentum}
\end{align}
Here $S_{\rm d}^{\sigma}$ is the disk four-momentum, $L_{\rm d}^{\mu\nu}$ is its orbital angular-momentum tensor. The covariant disk angular-momentum vector is then
\begin{equation}
 (J_{\rm d})_\rho
 =\frac{\epsilon_{\mu\nu\rho\sigma}
 L_{\rm d}^{\mu\nu}S_{\rm d}^{\sigma}}
 {2\sqrt{-S_{\rm d}^{\alpha}S_{{\rm d}\alpha}}}.
 \label{eq:disk-angular-momentum}
\end{equation}
Here $\epsilon_{\mu\nu\rho\sigma}$ is the Levi-Civita symbol. The vector $J_{\rm d}$ is constructed from the disk four-momentum $S_{\rm d}^{\sigma}$ and angular-momentum tensor $L_{\rm d}^{\mu\nu}$, and uses the same stress--energy tensor as the Teukolsky source in Equation~\eqref{eq:stress}. In the Cartesian frame whose $z$-axis is the BH spin axis, we define the disk tilt and precession phase by
\begin{align}
 \theta_{\rm d}
 &=\cos^{-1}\left(\frac{|J_{{\rm d},z}|}{|\boldsymbol J_{\rm d}|}\right), \\
 \phi_{\rm d}
 &=\tan^{-1}\left(J_{{\rm d},y},J_{{\rm d},x}\right).
 \label{eq:tilt-precession}
\end{align}
Here $\theta_{\rm d}$ is the tilt angle, $\phi_{\rm d}$ represents the direction of azimuthal precession.

To compare the relative contributions of azimuthal modes, we define
\begin{equation}
 E_m\equiv\int_{\widehat f>0}
 \frac{d\widehat E_m}{d\widehat f}\,d\widehat f,
 \label{eq:modal-energy}
\end{equation}
where $\widehat f=Mf$ is the dimensionless GW frequency.
To quantify the relative high-order contribution, we introduce the following ratio
\begin{equation}
 R_{\geq3}(m_{\max})\equiv
 \frac{\displaystyle\sum_{3\leq|m|\leq m_{\max}}E_m}
 {\displaystyle\sum_{|m|\leq2}E_m}.
 \label{eq:high-low-m}
\end{equation}
We adopt $m_{\max}=7$ and compare $m_{\max}=5$ and $6$ to show the convergence behavior. Each evolutionary sample
uses a $1200M$ source window with a cadence of $10M$.

\section{A multimode signature of disk tilt}
\label{sec:results}



\subsection{GW waveforms from aligned and tilted disks}
\label{sec:waveforms}
\begin{figure}[t]
\centering
\includegraphics[width=0.98\columnwidth]{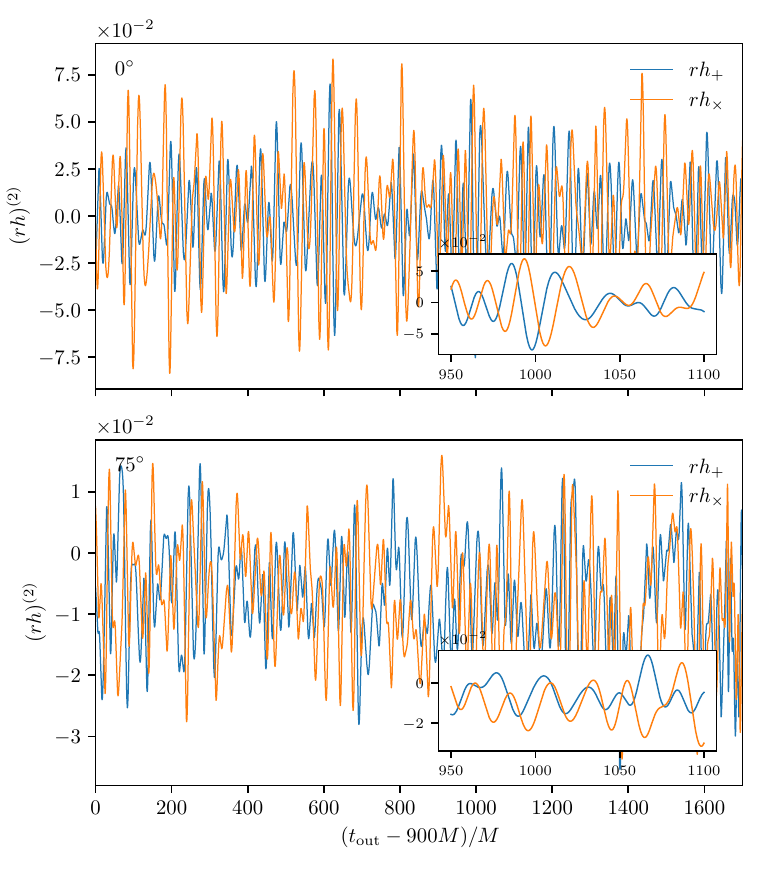}
\caption{The $m=2$ GW waveforms near $t=20000M$ for the initially
aligned (upper) and $75^\circ$ (lower) disks. Blue and orange curves denote
the plus and cross polarizations, respectively.}
\label{fig:waveform-m0}
\end{figure}

\citet{Yuan2025} found that a turbulent, aligned accretion disk produces a stochastic GW waveform. Its plus polarization mode, $h_+$, is about an order of magnitude stronger than its cross polarization mode $h_\times$, indicating that radial motion dominates GW emission.  We compare the $m=2$ waveforms of the aligned and initially $75^\circ$ disks near $t=2\times10^4M$ in Figure~\ref{fig:waveform-m0}. Both signals exhibit time-dependent oscillations in the plus and cross polarizations. Details of the waveform construction are given in Appendix~\ref{app:waveform-m0}.

\subsection{Alignment and redistribution of GW energy}
\label{sec:alignment}

\begin{figure*}[t]
\centering
\includegraphics[width=\textwidth]{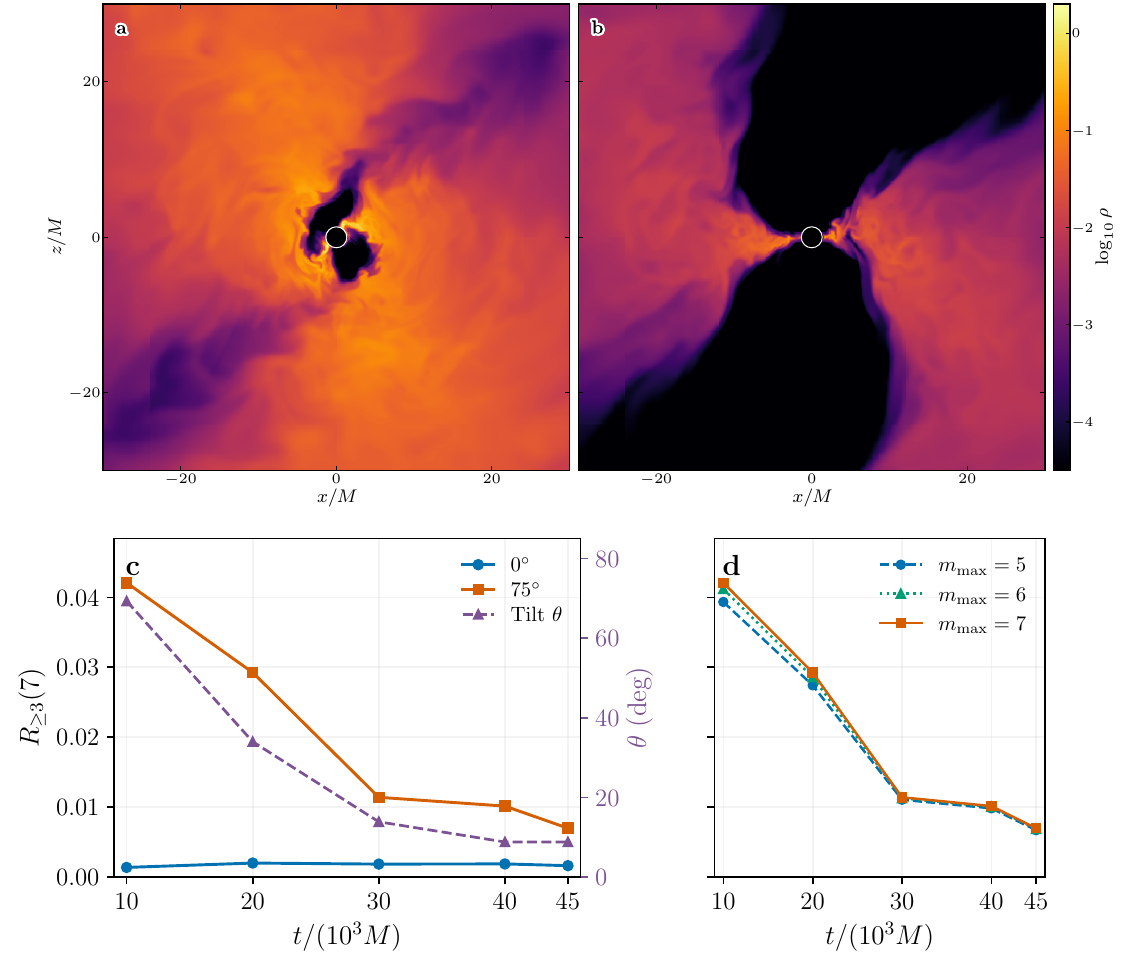}
\caption{Disk alignment and GW modal-energy evolution.
Panels (a) and (b) show density slices of the initially $75^\circ$ disk
at $t=12000M$ and $29000M$. Panel (c) compares $R_{\geq3}(7)$ for both
cases with the tilted disk's tilt angle $\theta$. Panel (d) demonstrates convergence of the tilted-disk ratio for
$m_{\max}=5$, 6, and 7.}
\label{fig:Rm}
\end{figure*}

Figure~\ref{fig:Rm} demonstrates the connection between disk geometry and the contribution of high-$|m|$ modes. Panel (a) shows the initially $75^\circ$ disk at $t=12000M$ where the dense flow crosses the inner region at a pronounced inclination to the BH equatorial plane. Panel (b) shows a more equatorial density distribution by $t=29000M$, with a low-density polar funnel along the spin axis. These snapshots illustrate the geometric reorientation.

Panel (c) presents the main result of this paper. It can be seen that the high-$|m|$ energy ratio decreases together with the tilt angle during the reorientation. Within the sampled interval, $10000M\leq t\leq30000M$, the tilt angle falls from $69.3^\circ$ to $13.9^\circ$, while $R_{\geq3}$ decreases from $0.042$ to $0.011$. The geometric reorientation is therefore accompanied by energy redistribution in the GW signals. At later epochs, the tilt approaches approximately $9^\circ$ and its evolution becomes much slower and $R_{\geq3}$ reaches $0.007$ at $t=45000M$. The ratio thus tracks the geometry of the disk in the strong-field dynamics. 

Panel (d) tests convergence with different $m_{\max}$. The ratios obtained with $m_{\max}=5$, 6, and 7 closely agree and all recover the same decline. This indicates that $|m|>7$ modes have negligible contribution to GW emission.

\section{Astrophysical implication}
\label{sec:scaling}

The linearity of the Teukolsky equation allows the radiation to be rescaled as $h_c\propto qM/D$ and $f\propto M^{-1}$ at fixed dimensionless disk structure, BH spin, and waveform duration, where $q\equiv M_{\rm disk}/M$ and $D$ is the luminosity distance. Adopting $M=5\times10^6M_\odot$, $D=4\,\mathrm{Mpc}$, and $q=0.05$, we obtain $h_c\simeq7.5\times10^{-21}$ for the initially tilted disk. The characteristic frequency is approximately $f\sim M^{-1}\sim3\,\mathrm{mHz}$. Such a source could be within reach of future space-based GW observatories, including LISA \citep{AmaroSeoane2017}, Taiji \citep{Luo:2021qji}, and TianQin \citep{TianQin:2020hid}.

\section{Discussion}
\label{sec:discussion}
In this paper, we perform GRMHD simulations of an initially tilted disk and compute its GW emission. By analyzing the GW signatures carried by different $m$ modes, we find that the energy distribution of $m$ modes tracks the tilt angle of the disk during the reorientation. Thus, GW signatures do provide geometric information of the black hole environment in the strong-field dynamics.

Here, we propose a simple geometrical model to explain the energy redistribution of the tilted disk. For a fixed orientation $(\theta_d,\phi_d)$, the rotation of the disk yields the transformation \citep{Boyle2013}
\begin{equation}
    h_{\ell m}^{\rm geom}(t)
    =
    e^{-im\phi_d}
    d^\ell_{m0}(\theta_d)\,
    h_{\ell 0}^{\rm d}(t),
    \label{eq:tilt_geometrical_mixing}
\end{equation}
where $d^\ell_{m0}$ is a Wigner small-$d$ matrix element and the upper index ${\rm d}$ denotes the disk-aligned frame. Since the rotation is time independent in this illustration, the fraction of the energy at fixed $\ell$ carried by each $m$ is $p_{\ell m}=|d^\ell_{m0}(\theta_d)|^2$. The corresponding rms azimuthal number satisfies
\begin{equation}
    m_{{\rm rms},\ell}^{\,2}
    \equiv
    \sum_{m=-\ell}^{\ell}m^2 p_{\ell m}
    =
    \frac{\ell(\ell+1)}{2}\sin^2\theta_d.
    \label{eq:tilt_azimuthal_width}
\end{equation}
In this idealized rotation model, a large tilt broadens the distribution over $m$. This provides a geometric interpretation of the declining curves in figure~\ref{fig:Rm}.

Our results illustrate how GW and electromagnetic observations can probe complementary properties of accretion flows. Despite GW emission, electromagnetic observations also depend on how the plasma emits. For example, electron thermodynamics and plasma composition can affect the predicted image, while the equation of state can alter the flow itself \citep{Palit2019,Ressler2015,Emami2021}. Together, GW and electromagnetic observations could constrain both the geometry of the disk and the radiative properties of the plasma.

\section{Conclusions}
\label{sec:conclusions}

We have combined three-dimensional GRMHD simulations with time-domain Teukolsky calculations to investigate GW emission from aligned and initially tilted turbulent disks. Our principal result is that the GW energy distribution over $m$ modes will track the decline of the tilt angle, indicating that GW signatures carry geometric information of the dynamic system in the strong-field. This suggest that future GW detection can be used to infer geometric information of the source.
Looking ahead, joint electromagnetic and GW observations could provide complementary constraints on accretion. Electromagnetic emission probes the radiating plasma and its microphysics, while the GW signatures encode the BH’s strong-field response and the large-scale geometry of the accretion flow.
\begin{acknowledgments}
The simulations were analyzed on the TDLI-Astro cluster at Shanghai Jiao Tong University.
\end{acknowledgments}

\appendix
\raggedbottom
\setcounter{equation}{0}
\renewcommand{\theequation}{A\arabic{equation}}
\renewcommand{\theHequation}{A.\arabic{equation}}
\setcounter{figure}{0}
\renewcommand{\thefigure}{A\arabic{figure}}
\renewcommand{\theHfigure}{A.\arabic{figure}}
\section[Near-horizon accretion rate and magnetic flux]{Near-horizon accretion rate\\ and magnetic flux}
\label{app:accretion-flux}

\label{app:numerical-setup}

Simulations in this work use a $30^3$ root grid with 10-levels of nested static mesh refinement centered on the black hole and inner disk, giving a finest cell width of $\Delta x=0.1\,r_g$.  We impose density and gas-pressure floors of $1\times 10^{-8}$ and $3.33\times10^{-11}$, respectively, and cap the Lorentz factor at $20$.

 We calculate the mass-accretion rate and normalized magnetic flux at the event horizon. Here $\rho$ is the rest-mass density, $u^r$ is the radial fluid four-velocity, $B^r$ is the radial magnetic-field component, and $dA$ is the  area element on the spherical surface:
\begin{align}
 \dot M_{\rm H}
  &=-\int\rho u^r\,dA, \label{eq:mdot-h}\\
 \Phi_{\rm H}
  &=\frac12\int |B^r|\,dA, \label{eq:phi-h}\\
 \varphi_{\rm H}
  &=\frac{\Phi_{\rm H}}{\sqrt{|\dot M_{\rm H}|}}. \label{eq:varphi-h}
\end{align}
 The factor of one half follows the horizon-flux normalization used for magnetized accretion flows \citep{Tchekhovskoy2011MAD,Porth2019,Jiang2025Retrograde}.

\begin{figure}[t]
\centering
\includegraphics[width=\columnwidth]{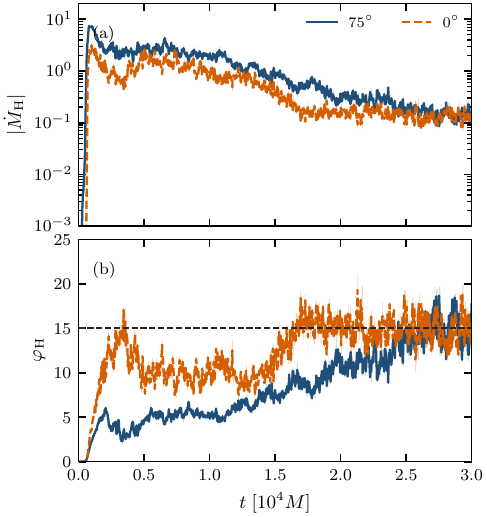}
\caption{Near-horizon accretion and magnetic flux of the
$0^\circ$ (orange dashed) and initially $75^\circ$ (blue solid) disks.
Panel (a) shows $|\dot M_{\rm H}|$ and panel (b) shows the
normalized flux $\varphi_{\rm H}=\Phi_{\rm H}/\sqrt{|\dot M_{\rm H}|}$, both
evaluated at the event horizon.    The dashed line in panel (b)
marks $\varphi_{\rm H}=15$. }
\label{fig:mdot-flux}
\end{figure}

Figure~\ref{fig:mdot-flux} shows the initial rise and subsequent decline of the accretion rate in both simulations. For the initially aligned case, the normalized magnetic flux is substantially higher than that in the tilted case during the first $2\times10^{4}\,M$, and reaches the MAD threshold at $\sim1.5\times10^{4}\,M$.
  In the initially tilted disk case, after
$t\sim2.5\times10^4M$, it reaches the threshold $\varphi_{\rm H}=15$ as well\citep{Tchekhovskoy2011MAD,McKinney2013}.  

\setcounter{equation}{0}
\renewcommand{\theequation}{B\arabic{equation}}
\renewcommand{\theHequation}{B.\arabic{equation}}
\setcounter{figure}{0}
\renewcommand{\thefigure}{B\arabic{figure}}
\renewcommand{\theHfigure}{B.\arabic{figure}}
\section{MRI resolution diagnostic}
\label{app:mri}

Following \citet{Porth2019} and the magnetorotational-instability resolution
criterion of \citet{Sorathia2012}, the directional quality factor is
\begin{align}
 Q_i&=\frac{\lambda_{{\rm MRI},i}}{\Delta x_i},\nonumber\\
 \lambda_{{\rm MRI},i}&=\frac{2\pi |v_{A,i}|}{|\Omega|},
 \label{eq:mri-quality}
\end{align}
Here $Q_i$ is the MRI quality factor (the number of cells per fastest-growing wavelength), $\lambda_{{\rm MRI},i}$ is that wavelength in direction $i$, and $\Delta x_i$ is the local grid-cell width in that direction.  The curves in Figure~\ref{fig:mri-quality} are density-weighted  averages over $2M<r<400\,r_g$, excluding floor density with $\rho<10^{-5}$.

\par\medskip
\noindent\begin{minipage}{\columnwidth}
\centering
\includegraphics[width=0.95\columnwidth,height=0.36\textheight,keepaspectratio]{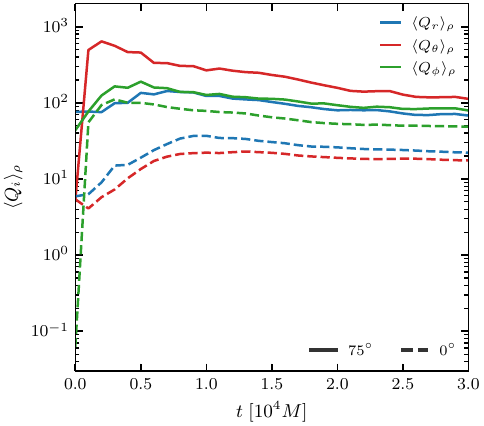}
\figcaption{Density-weighted MRI quality factors of the
$0^\circ$ disk (dashed) and initially $75^\circ$ disk (solid).
The blue, red, and green curves show $\langle Q_r\rangle_\rho$,
$\langle Q_\theta\rangle_\rho$, and $\langle Q_\phi\rangle_\rho$, averaged
over $2M<r<400\,r_g$ .\label{fig:mri-quality}}
\end{minipage}
\par\medskip

 In the late time, the density-weighted quality factors lie above the $Q_\theta\sim10$--$15$ and $Q_\phi\sim10$ benchmarks commonly used to judge whether the magnetorotational instability is resolved \citep{Sorathia2012,Porth2019}.

\setcounter{equation}{0}
\renewcommand{\theequation}{C\arabic{equation}}
\renewcommand{\theHequation}{C.\arabic{equation}}
\setcounter{figure}{0}
\renewcommand{\thefigure}{C\arabic{figure}}
\renewcommand{\theHfigure}{C.\arabic{figure}}
\section{\texorpdfstring{Time-domain waveforms and energy spectra}{Time-domain waveforms and energy spectra}}
\label{app:waveform-m0}

We extract the Teukolsky modes at the tortoise coordinate $r_*^{\rm ext}=500M$. Writing $\Psi_4=\sum_m e^{im\phi}\Psi_{4m}$, Equation~\eqref{eq:master} gives
\begin{equation}
 r\Psi_{4m}(t,r,\theta)
 =\left(1-\frac{ia\cos\theta}{r}\right)^{-4}
 \Phi_m(t,r_*(r),\theta).
 \label{eq:waveform-curvature-mode}
\end{equation}
For the waveforms in Figure~\ref{fig:waveform-m0}, we use the leading wave-zone approximation to the extraction at finite radius and integrate over the polar angle,
\begin{equation}
 \overline\Phi_m(t)=\int_0^\pi
 \Phi_m(t,r_*^{\rm ext},\theta)\sin\theta\,d\theta.
 \label{eq:waveform-polar-integral}
\end{equation}
With our convention $\Psi_4=(\ddot h_+-i\ddot h_\times)/2$, the strain is given by
\begin{equation}
 H_m(t)=2\int_0^t dt_1\int_0^{t_1}dt_2\,
 \overline\Phi_m(t_2).
 \label{eq:waveform-time-integral}
\end{equation}
\smallskip
The plus and cross waveforms in Figure~\ref{fig:waveform-m0} are $\operatorname{Re}H_2$ and $-\operatorname{Im}H_2$, respectively.

For the energy spectrum, we first define $\mathcal H_m(t,\theta)$ by
\begin{equation}
 r(h_+-ih_\times)=\sum_m e^{im\phi}\mathcal H_m(t,\theta),
 \qquad \ddot{\mathcal H}_m=2\Phi_m.
 \label{eq:angular-strain-modes}
\end{equation}
\smallskip
With $\widehat t=t/M$, $\widehat f=Mf$, $\widehat\Phi_m=M\Phi_m$, and $\widehat E_m=E_m^{\rm GW}/M$, we define
\begin{equation}
\widetilde{\widehat\Phi}_m(\widehat f,\theta)
 =\int_{-\infty}^{\infty}d\widehat t\,
 \widehat\Phi_m(\widehat t,\theta)e^{-2\pi i\widehat f\widehat t}.
 \label{eq:curvature-fourier-transform}
\end{equation}
Then, the one-sided GW energy spectrum can be expressed as
\begin{align}
 \frac{d\widehat E_m}{d\widehat f}
 ={}&\frac{1}{8\pi^2\widehat f^{\,2}}
 \int_0^\pi\sin\theta\,d\theta\,
 \Bigl[
 |\widetilde{\widehat\Phi}_m(\widehat f,\theta)|^2\nonumber\\
 &\hspace{4em} +
 |\widetilde{\widehat\Phi}_m(-\widehat f,\theta)|^2
 \Bigr].
 \label{eq:gw-modal-energy-spectrum}
\end{align}
Summing over $m$ gives the total energy spectrum.
For completeness, the characteristic strain is given by
\begin{equation}
 h_c^2(f)=\frac{2G}{\pi^2c^3D^2}
 \frac{dE^{\rm GW}}{df}.
 \label{eq:characteristic-strain-energy}
\end{equation}
Here $h_c^2=4f^2\langle|\widetilde h_+|^2+|\widetilde h_\times|^2\rangle_\Omega$, with $\langle X\rangle_\Omega=(4\pi)^{-1}\int X\,d\Omega$ and $D$ represents the distance between the source and the observer.

\bibliographystyle{aasjournalv7}
\bibliography{sample701}

@ARTICLE{Fishbone1976,
       author = {{Fishbone}, Leslie G. and {Moncrief}, Vincent},
        title = "{Relativistic fluid disks in orbit around Kerr black holes}",
      journal = {Astrophysical Journal},
         year = 1976,
        month = apr,
       volume = {207},
        pages = {962--976},
          doi = {10.1086/154565},
       adsurl = {https://ui.adsabs.harvard.edu/abs/1976ApJ...207..962F},
}

@ARTICLE{Fragile2005,
       author = {{Fragile}, P. Chris and {Anninos}, Peter},
        title = "{Hydrodynamic Simulations of Tilted Thick-Disk Accretion onto a Kerr Black Hole}",
      journal = {Astrophysical Journal},
         year = 2005,
        month = apr,
       volume = {623},
       number = {1},
        pages = {347--361},
          doi = {10.1086/428433},
       adsurl = {https://ui.adsabs.harvard.edu/abs/2005ApJ...623..347F},
}

@ARTICLE{Liska2018,
       author = {{Liska}, M. and {Hesp}, C. and {Tchekhovskoy}, A. and {Ingram}, A. and {van der Klis}, M. and {Markoff}, S.},
        title = "{Formation of precessing jets by tilted black hole discs in 3D general relativistic MHD simulations}",
      journal = {Monthly Notices of the Royal Astronomical Society},
         year = 2018,
        month = feb,
       volume = {474},
       number = {1},
        pages = {L81--L85},
          doi = {10.1093/mnrasl/slx174},
       adsurl = {https://ui.adsabs.harvard.edu/abs/2018MNRAS.474L..81L},
}

@ARTICLE{White2019,
       author = {{White}, Christopher J. and {Quataert}, Eliot and {Blaes}, Omer},
        title = "{Tilted Disks around Black Holes: A Numerical Parameter Survey for Spin and Inclination Angle}",
      journal = {Astrophysical Journal},
         year = 2019,
        month = jun,
       volume = {878},
       number = {1},
          eid = {51},
        pages = {51},
          doi = {10.3847/1538-4357/ab089e},
archivePrefix = {arXiv},
       eprint = {1902.09662},
 primaryClass = {astro-ph.HE},
       adsurl = {https://ui.adsabs.harvard.edu/abs/2019ApJ...878...51W},
}

@ARTICLE{Jiang2025Retrograde,
       author = {{Jiang}, Hong-Xuan and {Mizuno}, Yosuke and {Lai}, Dong and {Dihingia}, Indu K. and {Fromm}, Christian M.},
        title = "{Magnetically Driven Retrograde Precession in Misaligned Black Hole Accretion Flows}",
      journal = {Astrophysical Journal},
         year = 2025,
        month = dec,
       volume = {995},
       number = {2},
          eid = {112},
        pages = {112},
          doi = {10.3847/1538-4357/ae19ee},
       adsurl = {https://ui.adsabs.harvard.edu/abs/2025ApJ...995..112J},
}

@ARTICLE{Tchekhovskoy2011MAD,
       author = {{Tchekhovskoy}, Alexander and {Narayan}, Ramesh and {McKinney}, Jonathan C.},
        title = "{Efficient generation of jets from magnetically arrested accretion on a rapidly spinning black hole}",
      journal = {Monthly Notices of the Royal Astronomical Society},
         year = 2011,
        month = nov,
       volume = {418},
       number = {1},
        pages = {L79--L83},
          doi = {10.1111/j.1745-3933.2011.01147.x},
archivePrefix = {arXiv},
       eprint = {1108.0412},
 primaryClass = {astro-ph.HE},
       adsurl = {https://ui.adsabs.harvard.edu/abs/2011MNRAS.418L..79T},
}

@ARTICLE{Stone2020AthenaPP,
       author = {{Stone}, James M. and {Tomida}, Kengo and {White}, Christopher J. and {Felker}, Kyle G.},
        title = "{The Athena++ Adaptive Mesh Refinement Framework: Design and Magnetohydrodynamic Solvers}",
      journal = {Astrophysical Journal Supplement Series},
         year = 2020,
        month = mar,
       volume = {249},
       number = {1},
          eid = {4},
        pages = {4},
          doi = {10.3847/1538-4365/ab929b},
archivePrefix = {arXiv},
       eprint = {2005.06651},
 primaryClass = {astro-ph.IM},
       adsurl = {https://ui.adsabs.harvard.edu/abs/2020ApJS..249....4S},
}

@ARTICLE{Stone2024AthenaK,
       author = {{Stone}, James M. and {Mullen}, Patrick D. and {Fielding}, Drummond and {Grete}, Philipp and {Guo}, Minghao and {Kempski}, Philipp and {Most}, Elias R. and {White}, Christopher J. and {Wong}, George N.},
        title = "{AthenaK: A Performance-Portable Version of the Athena++ AMR Framework}",
      journal = {arXiv e-prints},
         year = 2024,
        month = sep,
          eid = {arXiv:2409.16053},
        pages = {arXiv:2409.16053},
          doi = {10.48550/arXiv.2409.16053},
archivePrefix = {arXiv},
       eprint = {2409.16053},
 primaryClass = {astro-ph.IM},
       adsurl = {https://ui.adsabs.harvard.edu/abs/2024arXiv240916053S},
}

@ARTICLE{Yuan2025,
       author = {{Yuan}, Chen and {Cardoso}, Vitor and {Duque}, Francisco and {Younsi}, Ziri},
        title = "{Gravitational waves from accretion disks: Turbulence, mode excitation, and prospects for future detectors}",
      journal = {Physical Review D},
         year = 2025,
        month = mar,
       volume = {111},
       number = {6},
          eid = {063048},
        pages = {063048},
          doi = {10.1103/PhysRevD.111.063048},
       adsurl = {https://ui.adsabs.harvard.edu/abs/2025PhRvD.111f3048Y},
}

@ARTICLE{ArayaGochez2004,
       author = {{Araya-Gochez}, Rafael Angel},
        title = "{Gravitational Waves from Hyper-Accretion onto Nascent Black Holes}",
      journal = {Monthly Notices of the Royal Astronomical Society},
         year = 2004,
       volume = {355},
        pages = {336},
          doi = {10.1111/j.1365-2966.2004.07851.x},
archivePrefix = {arXiv},
       eprint = {astro-ph/0311001},
}

@ARTICLE{Wessel2021,
       author = {{Wessel}, Erik and {Paschalidis}, Vasileios and {Tsokaros}, Antonios and {Ruiz}, Milton and {Shapiro}, Stuart L.},
        title = "{Gravitational Waves from Disks Around Spinning Black Holes: Simulations in Full General Relativity}",
      journal = {Physical Review D},
         year = 2021,
       volume = {103},
        pages = {043013},
          doi = {10.1103/PhysRevD.103.043013},
archivePrefix = {arXiv},
       eprint = {2011.04077},
}

@ARTICLE{Gottlieb2024,
       author = {{Gottlieb}, Ore and {Levinson}, Amir and {Levin}, Yuri},
        title = "{In LIGO's Sight? Vigorous Coherent Gravitational Waves from Cooled Collapsar Disks}",
      journal = {The Astrophysical Journal Letters},
         year = 2024,
       volume = {972},
        pages = {L4},
          doi = {10.3847/2041-8213/ad697c},
archivePrefix = {arXiv},
       eprint = {2406.19452},
}

@ARTICLE{Ruiz2026,
       author = {{Ruiz}, Milton and {Tsokaros}, Antonios and {Shapiro}, Stuart L.},
        title = "{Multimessenger Signatures of Tilted, Self-Gravitating, Black Hole Disks}",
      journal = {Physical Review Letters},
         year = 2026,
       volume = {137},
       number = {5},
        pages = {051402},
          doi = {10.1103/7zd8-yygd},
archivePrefix = {arXiv},
       eprint = {2604.03368},
 primaryClass = {gr-qc},
}

@ARTICLE{Teukolsky1973,
       author = {{Teukolsky}, Saul A.},
        title = "{Perturbations of a Rotating Black Hole. I. Fundamental Equations for Gravitational, Electromagnetic, and Neutrino-Field Perturbations}",
      journal = {Astrophysical Journal},
         year = 1973,
        month = sep,
       volume = {185},
        pages = {635--648},
          doi = {10.1086/152444},
       adsurl = {https://ui.adsabs.harvard.edu/abs/1973ApJ...185..635T},
}

@ARTICLE{Bardeen1975,
       author = {{Bardeen}, James M. and {Petterson}, Jacobus A.},
        title = "{The Lense-Thirring Effect and Accretion Disks around Kerr Black Holes}",
      journal = {Astrophysical Journal Letters},
         year = 1975,
        month = jan,
       volume = {195},
        pages = {L65},
          doi = {10.1086/181711},
       adsurl = {https://ui.adsabs.harvard.edu/abs/1975ApJ...195L..65B},
}

@ARTICLE{Chatterjee2025,
       author = {{Chatterjee}, Koushik and {Naaz}, Sajal and {Liska}, Matthew and {Tchekhovskoy}, Alexander and {Markoff}, Sera},
        title = "{Misaligned magnetized accretion flows onto spinning black holes: Magneto-spin alignment, outflow power, and intermittency}",
      journal = {Physical Review D},
         year = 2025,
       volume = {112},
       number = {6},
          eid = {063013},
        pages = {063013},
          doi = {10.1103/PhysRevD.112.063013},
archivePrefix = {arXiv},
       eprint = {2311.00432},
 primaryClass = {astro-ph.HE},
       adsurl = {https://ui.adsabs.harvard.edu/abs/2025PhRvD.112f3013C},
}

@ARTICLE{AmaroSeoane2017,
       author = {{Amaro-Seoane}, Pau and others},
        title = "{Laser Interferometer Space Antenna}",
      journal = {arXiv e-prints},
         year = 2017,
        month = feb,
          eid = {arXiv:1702.00786},
        pages = {arXiv:1702.00786},
archivePrefix = {arXiv},
       eprint = {1702.00786},
 primaryClass = {astro-ph.IM},
       adsurl = {https://ui.adsabs.harvard.edu/abs/2017arXiv170200786A},
}

@ARTICLE{Abbott2016GW150914,
       author = {{Abbott}, B.~P. and others},
        title = "{Observation of Gravitational Waves from a Binary Black Hole Merger}",
      journal = {Physical Review Letters},
         year = 2016,
        month = feb,
       volume = {116},
       number = {6},
          eid = {061102},
        pages = {061102},
          doi = {10.1103/PhysRevLett.116.061102},
archivePrefix = {arXiv},
       eprint = {1602.03837},
 primaryClass = {gr-qc},
       adsurl = {https://ui.adsabs.harvard.edu/abs/2016PhRvL.116f1102A},
}

@ARTICLE{Abbott2017BNS,
       author = {{Abbott}, B.~P. and others},
        title = "{GW170817: Observation of Gravitational Waves from a Binary Neutron Star Inspiral}",
      journal = {Physical Review Letters},
         year = 2017,
        month = oct,
       volume = {119},
       number = {16},
          eid = {161101},
        pages = {161101},
          doi = {10.1103/PhysRevLett.119.161101},
archivePrefix = {arXiv},
       eprint = {1710.05832},
 primaryClass = {gr-qc},
       adsurl = {https://ui.adsabs.harvard.edu/abs/2017PhRvL.119p1101A},
}

@ARTICLE{BlandfordZnajek1977,
       author = {{Blandford}, Roger D. and {Znajek}, Roman L.},
        title = "{Electromagnetic extraction of energy from Kerr black holes}",
      journal = {Monthly Notices of the Royal Astronomical Society},
         year = 1977,
        month = may,
       volume = {179},
       number = {3},
        pages = {433--456},
          doi = {10.1093/mnras/179.3.433},
       adsurl = {https://ui.adsabs.harvard.edu/abs/1977MNRAS.179..433B},
}

@ARTICLE{Greenhill2003Circinus,
       author = {{Greenhill}, Lincoln J. and others},
        title = "{A Warped Accretion Disk and Wide-Angle Outflow in the Inner Parsec of the Circinus Galaxy}",
      journal = {Astrophysical Journal},
         year = 2003,
        month = jun,
       volume = {590},
       number = {1},
        pages = {162--173},
          doi = {10.1086/374870},
       adsurl = {https://ui.adsabs.harvard.edu/abs/2003ApJ...590..162G},
}

@ARTICLE{Kosec2023HerX1,
       author = {{Kosec}, P. and others},
        title = "{A precessing accretion disc in the X-ray binary Her X-1}",
      journal = {Nature Astronomy},
         year = 2023,
       volume = {7},
        pages = {715--723},
          doi = {10.1038/s41550-023-01931-7},
       adsurl = {https://ui.adsabs.harvard.edu/abs/2023NatAs...7..715K},
}

@ARTICLE{King2005,
       author = {{King}, A.~R. and {Lubow}, S.~H. and {Ogilvie}, G.~I. and {Pringle}, J.~E.},
        title = "{Aligning spinning black holes and accretion discs}",
      journal = {Monthly Notices of the Royal Astronomical Society},
         year = 2005,
        month = oct,
       volume = {363},
       number = {1},
        pages = {49--56},
          doi = {10.1111/j.1365-2966.2005.09378.x},
       adsurl = {https://ui.adsabs.harvard.edu/abs/2005MNRAS.363...49K},
}

@ARTICLE{KingPringle2006,
       author = {{King}, A.~R. and {Pringle}, J.~E.},
        title = "{Growing supermassive black holes by chaotic accretion}",
      journal = {Monthly Notices of the Royal Astronomical Society},
         year = 2006,
        month = jun,
       volume = {373},
       number = {1},
        pages = {L90--L92},
          doi = {10.1111/j.1745-3933.2006.00180.x},
       adsurl = {https://ui.adsabs.harvard.edu/abs/2006MNRAS.373L..90K},
}

@ARTICLE{Nixon2012,
       author = {{Nixon}, Chris and {King}, Andrew and {Price}, Daniel and {Frank}, Juhan},
        title = "{Tearing up the Disk: How Black Holes Accrete}",
      journal = {Astrophysical Journal Letters},
         year = 2012,
        month = oct,
       volume = {757},
       number = {2},
          eid = {L24},
        pages = {L24},
          doi = {10.1088/2041-8205/757/2/L24},
       adsurl = {https://ui.adsabs.harvard.edu/abs/2012ApJ...757L..24N},
}

@ARTICLE{ScheuerFeiler1996,
       author = {{Scheuer}, P.~A.~G. and {Feiler}, R.},
        title = "{The realignment of a black hole misaligned with its accretion disc}",
      journal = {Monthly Notices of the Royal Astronomical Society},
         year = 1996,
        month = sep,
       volume = {282},
       number = {1},
        pages = {291--294},
          doi = {10.1093/mnras/282.1.291},
       adsurl = {https://ui.adsabs.harvard.edu/abs/1996MNRAS.282..291S},
}

@ARTICLE{Nealon2015,
       author = {{Nealon}, Rebecca and {Price}, Daniel J. and {Nixon}, Chris J.},
        title = "{On the Bardeen-Petterson effect in black hole accretion discs}",
      journal = {Monthly Notices of the Royal Astronomical Society},
         year = 2015,
        month = apr,
       volume = {448},
       number = {2},
        pages = {1526--1540},
          doi = {10.1093/mnras/stv014},
archivePrefix = {arXiv},
       eprint = {1501.01687},
 primaryClass = {astro-ph.HE},
       adsurl = {https://ui.adsabs.harvard.edu/abs/2015MNRAS.448.1526N},
}

@ARTICLE{Porth2019,
       author = {{Porth}, Oliver and {Chatterjee}, Koushik and {Narayan}, Ramesh and {Gammie}, Charles F. and {Mizuno}, Yosuke and {Anninos}, Peter and {Baker}, John G. and {Bugli}, Matteo and {Chan}, Chi-kwan and {Davelaar}, Jordy and {Del Zanna}, Luca and {Etienne}, Zachariah B. and {Farris}, Brian D. and {Kelly}, Bernard J. and {Liska}, Matthew and {Markoff}, Sera and {McKinney}, Jonathan C. and {Moscibrodzka}, Monika and {Noble}, Scott C. and {Olivares}, Hector and {Prather}, Ben and {Rezzolla}, Luciano and {Ryan}, Benjamin R. and {Stone}, James M. and {Tchekhovskoy}, Alexander and {Toma}, Kenji and {White}, Christopher J. and {Younsi}, Ziri and {Zanotti}, Olindo},
        title = "{The Event Horizon General Relativistic Magnetohydrodynamic Code Comparison Project}",
      journal = {Astrophysical Journal Supplement Series},
         year = 2019,
        month = aug,
       volume = {243},
       number = {2},
          eid = {26},
        pages = {26},
          doi = {10.3847/1538-4365/ab29fd},
archivePrefix = {arXiv},
       eprint = {1904.04923},
 primaryClass = {astro-ph.HE},
       adsurl = {https://ui.adsabs.harvard.edu/abs/2019ApJS..243...26P},
}

@ARTICLE{Sorathia2012,
       author = {{Sorathia}, Kareem A. and {Reynolds}, Christopher S. and {Stone}, James M. and {Beckwith}, Kris},
        title = "{The Global Structure of Magnetorotational Turbulence in Accretion Disks}",
      journal = {Astrophysical Journal},
         year = 2012,
        month = apr,
       volume = {749},
       number = {2},
          eid = {189},
        pages = {189},
          doi = {10.1088/0004-637X/749/2/189},
archivePrefix = {arXiv},
       eprint = {1111.0962},
 primaryClass = {astro-ph.HE},
       adsurl = {https://ui.adsabs.harvard.edu/abs/2012ApJ...749..189S},
}

@ARTICLE{McKinney2013,
       author = {{McKinney}, Jonathan C. and {Tchekhovskoy}, Alexander and {Blandford}, Roger D.},
        title = "{Alignment of Magnetized Accretion Disks and Relativistic Jets with Spinning Black Holes}",
      journal = {Science},
         year = 2013,
        month = jan,
       volume = {339},
       number = {6115},
        pages = {49--52},
          doi = {10.1126/science.1230811},
archivePrefix = {arXiv},
       eprint = {1211.3651},
 primaryClass = {astro-ph.CO},
       adsurl = {https://ui.adsabs.harvard.edu/abs/2013Sci...339...49M},
}

@article{Boyle2013,
  author        = {Boyle, Michael},
  title         = {Angular velocity of gravitational radiation from
                   precessing binaries and the corotating frame},
  journal       = {Phys. Rev. D},
  volume        = {87},
  pages         = {104006},
  year          = {2013},
  doi           = {10.1103/PhysRevD.87.104006},
  eprint        = {1302.2919},
  archivePrefix = {arXiv},
  primaryClass  = {gr-qc}
}

@article{Palit2019,
  author = {Palit, Ishika and Janiuk, Agnieszka and Sukova, Petra},
  title = {Effects of adiabatic index on the sonic surface and time variability of low angular momentum accretion flows},
  journal = {Monthly Notices of the Royal Astronomical Society},
  volume = {487},
  number = {1},
  pages = {755--768},
  year = {2019},
  doi = {10.1093/mnras/stz1296}
}

@article{Ressler2015,
  author = {Ressler, Sean M. and Tchekhovskoy, Alexander and Quataert, Eliot and Chandra, Mani and Gammie, Charles F.},
  title = {Electron thermodynamics in {GRMHD} simulations of low-luminosity black hole accretion},
  journal = {Monthly Notices of the Royal Astronomical Society},
  volume = {454},
  number = {2},
  pages = {1848--1870},
  year = {2015},
  doi = {10.1093/mnras/stv2084},
  eprint = {1509.04717},
  archivePrefix = {arXiv}
}

@article{Emami2021,
  author = {Emami, Razieh and Anantua, Richard and Chael, Andrew A. and Loeb, Abraham},
  title = {Positron effects on polarized images and spectra from jet and accretion flow models of {M87*} and {Sgr A*}},
  journal = {The Astrophysical Journal},
  volume = {923},
  number = {2},
  pages = {272},
  year = {2021},
  doi = {10.3847/1538-4357/ac2950},
  eprint = {2101.05327},
  archivePrefix = {arXiv}
}

@article{LVK2026GW250114,
  author  = {{The LIGO Scientific Collaboration} and
             {The Virgo Collaboration} and
             {The KAGRA Collaboration}},
  title   = {Black Hole Spectroscopy and Tests of General Relativity with {GW250114}},
  journal = {Physical Review Letters},
  volume  = {136},
  number  = {4},
  pages   = {041403},
  year    = {2026},
  doi     = {10.1103/6c61-fm1n},
  eprint  = {2509.08099},
  archivePrefix = {arXiv},
  primaryClass  = {gr-qc}
}

@article{Wang:2025oaq,
    author = "Wang, Yanan and others",
    title = "{Detection of disk-jet coprecession in a tidal disruption event}",
    eprint = "2511.12477",
    archivePrefix = "arXiv",
    primaryClass = "astro-ph.HE",
    doi = "10.1126/sciadv.ady9068",
    journal = "Sci. Adv.",
    volume = "11",
    number = "50",
    pages = "ady9068",
    year = "2025"
}

@article{Fernandez:2025qbq,
    author = "Fern{\'a}ndez, Rodrigo and Janke, Silas and Dean, Coleman and Tamborra, Irene",
    title = "{Collapsar disk outflows: Detectable neutrino and gravitational wave signatures}",
    eprint = "2507.17836",
    archivePrefix = "arXiv",
    primaryClass = "astro-ph.HE",
    doi = "10.1103/d1q5-2qxx",
    journal = "Phys. Rev. D",
    volume = "112",
    number = "12",
    pages = "123047",
    year = "2025"
}

@ARTICLE{maccarone2002,
  author = {{Maccarone}, T. J.},
  title = "{On the Misalignment of Jets in Microquasars}",
  journal = {\mnras},
  year = 2002,
  volume = {336},
  number = {4},
  pages = {1371--1376},
  doi = {10.1046/j.1365-8711.2002.05876.x}
}

@ARTICLE{balbus1991,
  author = {{Balbus}, Steven A. and {Hawley}, John F.},
  title = "{A Powerful Local Shear Instability in Weakly Magnetized Disks.
            I. Linear Analysis}",
  journal = {\apj},
  year = 1991,
  volume = {376},
  pages = {214--233},
  doi = {10.1086/170270}
}

@ARTICLE{narayan2012,
  author = {{Narayan}, R. and {Sadowski}, A. and {Penna}, R. F. and
            {Kulkarni}, A. K.},
  title = "{GRMHD Simulations of Magnetized Advection-dominated Accretion on
            a Non-spinning Black Hole: Role of Outflows}",
  journal = {\mnras},
  year = 2012,
  volume = {426},
  number = {4},
  pages = {3241--3259},
  doi = {10.1111/j.1365-2966.2012.22002.x},
  eprint = {1206.1213},
  archivePrefix = {arXiv},
  primaryClass = {astro-ph.HE}
}

@article{TianQin:2020hid,
    author = "Mei, Jianwei and others",
    collaboration = "TianQin",
    title = "{The TianQin project: current progress on science and technology}",
    eprint = "2008.10332",
    archivePrefix = "arXiv",
    primaryClass = "gr-qc",
    doi = "10.1093/ptep/ptaa114",
    journal = "PTEP",
    volume = "2021",
    number = "5",
    pages = "05A107",
    year = "2021",
    note = "[Erratum: PTEP 2021, 059201 (2021)]"
}

@article{Luo:2021qji,
    author = "Luo, Ziren and Wang, Yan and Wu, Yueliang and Hu, Wenrui and Jin, Gang",
    title = "{The Taiji program: A concise overview}",
    doi = "10.1093/ptep/ptaa083",
    journal = "PTEP",
    volume = "2021",
    number = "5",
    pages = "05A108",
    year = "2021"
}

\end{document}